\documentclass[12pt]{article}

\input amssym.def

\input amssym

\newcommand{\be}{\begin{equation}}
\newcommand{\ee}{\end{equation}}

\title{A spherically symmetric dust-space-time
with a NUT-like rotation}

\date{}

\author{H\'ector Vargas-Rodr{\'\i}guez
\footnote{Physics Department, C.U.C.E.I., University of
Guadalajara; and Mathematics and Physics Department, I.T.E.S.O.,
Guadalajara, Mexico.  E-mail: hvargas@iteso.mx
 }}

\begin{document}

\maketitle

\section{Introduction}
In this paper we present a deduction of a solution (already discovered
in 1983 by Luk\'acs {\em et al}) of the
Einstein's equations, employing the Mitskievich's field theoretic
description of perfect fluids. This solution describes a
dust-space-time with a spherical-like symmetry and a NUT-like
rotation. This solution is of Petrov type $D$, and has an isometry
group $G_4$. It also admits closed timelike geodesics. It has
Minkowski space-time as a limit, when both dust and rotation
disappear.

This paper is organized as follows, in Section two the
Mitskievich's field theoretic description of perfect fluids is
presented, then in Section three we show this solution and its
properties.

Below we are working in four space-time dimensions with signature
\linebreak $(+,-,-,-)$, Greek indices being four-dimensional. The
Ricci tensor is $R_{\mu\nu}={R^\alpha}_{\mu\nu\alpha}$, thus
Einstein's equations take the form
$R_{\mu\nu}-\frac{1}{2}Rg_{\mu\nu}= -\varkappa T_{\mu\nu}$.

\section{Rotating fluids in field-theoretic-description}

\vspace*{0.5cm}

\noindent Perfect fluids can be conveniently described with use of
the Lagrangian formalism, especially in the absence of rotation
[2, 3]. In this case they are represented via the 2-form field
potential $B=\frac{1}{2!}B_{\mu\nu}dx^\mu\wedge dx^\nu$, the
respective field intensity being $G=dB=\frac{1}{2}
B_{\mu\nu;\lambda} dx^\lambda\wedge dx^\mu\wedge dx^\nu$
($B_{[\mu\nu;\lambda]} \equiv B_{[\mu\nu,\lambda]}$, and
$G_{\lambda\mu\nu}=B_{\lambda\mu,\nu}+B_{\mu\nu,\lambda}+
B_{\nu\lambda,\mu}$) whose invariant $J=\ast(G\wedge\ast G)$ (we
shall also denote $\ast G= \tilde{G}$) is used in constructing the
fluid Lagrangian density ${\cal L}$ $=\sqrt{-g}L(J)$. Here the
Hodge star $\ast$ denotes, as usual, a generalization of the dual
conjugation applied to Cartan exterior forms: with an $r$-form
$\alpha=\alpha_{\nu_1\dots\nu_r} dx^{\nu_1}\wedge\cdots\wedge
dx^{\nu_r}$, it yields a $(4-r)$-form $\ast\alpha$ with the
components $(\ast\alpha)_{\nu_1\dots \nu_{4-r}}=\frac{1}{r!}
E_{\nu_1\dots\nu_{4-r}\nu_{5-r} \dots\nu_4}\alpha^{\nu_{5-r}
\dots\nu_4}$ where $E_{\varkappa \lambda\mu\nu}= \sqrt{-g}
\epsilon_{\varkappa \lambda\mu\nu}$ and $E^{\varkappa\lambda\mu
\nu}=-(1/\sqrt{-g}) \epsilon_{\varkappa \lambda\mu\nu}$ are co-
and contravariant axial skew rank-4 tensors, $\epsilon_{\varkappa
\lambda\mu\nu}= \epsilon_{[ \varkappa\lambda\mu\nu]}$,
$\epsilon_{0123}=+1$ being the Levi-Civit\`a symbol ({\it cf.}
somewhat other notations in [1]).

The reason why this description of perfect fluids is valid, is
simply the fact that the stress-energy tensor of a 2-form field is
\be
T^\beta_\alpha=2J\frac{dL}{dJ}b^\beta_\alpha
-L\delta^\beta_\alpha \label{1}
\ee
where
\be
b^\beta_\alpha=\delta^\beta_\alpha- u_\alpha u^\beta,
b^\beta_\alpha u^\alpha=0= b^\beta_\alpha u_\beta, ~
u=J^{-1/2}\tilde{G}.\label{2}
\ee
When $u$ is timelike ($u\cdot
u=+1$, as we above supposed it to be), we come to the usual
perfect fluid whose (arbitrary) equation of state is determined by
the dependence of $L$ on its only argument, $J$ (see [2], [3])
[however when it is spacelike, the `fluid' is tachyonic (see for
some details [4], subsection 3.2)]. Since $b^\beta_\alpha$ is the
projector on the (local) subspace orthogonal to the congruence of
$u$, the latter is an eigenvector of the stress-energy tensor with
the eigenvalue $(-L)$, $T^\beta_\alpha u^\alpha=-Lu^\beta$, while
any vector orthogonal to $u$ is also eigenvector, now with the
three-fold eigenvalue $2J\frac{dL}{dJ} -L$. This is the property
of the stress-energy tensor of a perfect fluid possessing the
proper mass density $\mu$ and pressure $p$ (in its local rest
frame):
\be
\label{mu_p} \mu=-L, ~ ~ p=L-2J\frac{dL}{dJ}.
\ee
Below we consider perfect fluids characterized by the simplest
equation of state \be \label{eqstate} p=(2k-1)\mu \ee (the
frequently used notation is $2k=\gamma$) which correspond to the
Lagrangian $L=-\sigma|J|^k$, $\sigma>0$. In a four-dimensional
spacetime, the important special cases are: the incoherent dust
($p=0$) for $k=1/2$, intrinsically relativistic incoherent
radiation ($p=\mu/3$) for $k=2/3$, and hyperrelativistic stiff
matter ($p=\mu$) for $k=1$.

However the 2-form field equation which follows from the above
Lagrangian,
\be
\label{3} \left(\sqrt{-g}\frac{dL}{dJ}
G^{\lambda\mu\nu}\right)_{,\nu}=0 ~ \Longleftrightarrow ~
d\left(J^{1/2}\frac{dL}{dJ}u\right)=0,
\ee
only means that the
$\tilde{G}$ (equivalently, $u$) congruence is non-rotating. To
describe a rotating fluid, one has to introduce in (\ref{3}) a
non-zero right-hand side. This, in a sharp contrast to the usual
equations of mathematical physics ({\it cf.}, for example,
electrodynamics), {\em cannot then be interpreted as a usual
source term} (this was stressed in [4]): its meaning essentially
is to indicate the presence of rotation ($u\wedge du\neq 0$). To
this end it is necessary to consider one more field which we call
the Machian one, a 3-form field $C$ with the intensity $W=dC$ (see
[2, 3]). In terms of $L(K)$, $K=-(1/4!)W_{\kappa\lambda\mu\nu}
W^{\kappa\lambda\mu\nu}=\tilde{W}^2$, its equations reduce to
\be\label{4} \left(\sqrt{-g}\frac{dL}{dK}
W^{\kappa\lambda\mu\nu}\right)_{,\nu}=0 ~ \Rightarrow ~
K^{1/2}\frac{dL}{dK}=\mbox{const.} \ee We use also the duality
relations $B\!\stackrel{\mu\nu}{*}=\frac{1}{2}
E^{\mu\nu\alpha\beta} B_{\alpha\beta}$, $G_{\lambda\mu\nu}=
\tilde{G}^\kappa E_{\kappa\lambda\mu\nu}$, $W_{\kappa\lambda
\mu\nu}=\tilde{W}E_{\kappa\lambda\mu\nu}$. Moreover,
${B\!\stackrel{\mu\nu}{*}}_{;\nu}\equiv -(\ast G)^\mu$.

Since we were confronted with the no rotation property of perfect
fluid when the rank 2 field was considered to be free, the only
remedy now is to introduce a non-trivial ``source'' term in the
$r=2$ field equations, thus to consider the non-free field case
or, at least, to include in the Lagrangian a dependence on the
rank 2 field potential $B$. The simplest way to do this is to
introduce in the Lagrangian density dependence on a new invariant
$J_1= -B_{[\kappa\lambda}B_{\mu\nu]}B^{[\kappa\lambda}B^{\mu\nu]}$
which does not spoil the structure of stress-energy tensor,
simultaneously yielding a ``source'' term (thus permitting to
destroy the no rotation property) without changing the divergence
term in the $r=2$ field equations. We shall use below three
invariants: the obvious ones, $J$ and $K$, and the just introduced
invariant of the $r=2$ field {\em potential}, $J_1$. Then \be
\label{6} B_{[\kappa\lambda}B_{\mu\nu]}=-\frac{2}{4!}
B_{\alpha\beta} B\!\stackrel{\alpha\beta}{\ast}E_{\kappa\lambda
\mu\nu}. \ee Thus $J_1^{1/2}= 6^{-1/2} B_{\alpha\beta}
B\!\stackrel{\alpha\beta} {\ast}$. In fact, $J_1=0$, if $B$ is a
simple bivector ($B=a\wedge b$, $a$ and $b$ being 1-forms). This
{\em cannot however annul} the expression which this invariant
contributes to the $r=2$ field equations: up to a factor, it is
equal to $\partial J_1^{1/2}/\partial B_{\mu\nu}\neq 0$. Thus let
the Lagrangian density be \be \label{7} {\cal
L}=\sqrt{-g}\left(L(J)+M(K) J_1^{1/2}\right), \ee so that the
$r=2$ field equations take the form ({\em cf.} (\ref{3})) \be
\label{x} d\left(\frac{dL}{dJ} \tilde{G}\right)=
\sqrt{\frac{2}{3}}M(K)B ~ \Leftrightarrow ~
\left(\sqrt{-g}\frac{dL}{dJ}{G^{\alpha\beta\nu}}\right)_{,\nu}=
\sqrt{-g} \sqrt{\frac{2}{3}}M(K)B\!\stackrel{\alpha\beta}{\ast}.
\ee In their turn, the $r=3$ field equations ({\em cf.} (\ref{4}))
yield the first integral \be\label{y} J_1^{1/2}
K^{1/2}\frac{dM}{dK}= \mbox{const}\equiv 0 \ee (since $J_1=0$). We
know from [1, 2] that $K$ (hence, $M$) {\em arbitrarily} depends
on the space-time coordinates, if only the $r=3$ field equations
are taken into account, and the Machian field $K$ has to be
essentially non-constant.

The stress-energy tensor which corresponds to the new Lagrangian
density (\ref{7}), automatically coincides with its previous form
(\ref{1}), since $J_1=0$. For a perfect fluid with the equation of
state $p=(2k-1)\mu$, one finds $L=-\sigma J^k$, thus
$T^\beta_\alpha=-2kLu_\alpha u^\beta+(2k-1)L \delta^\beta_\alpha$.
Then the traditional perfect fluid language is obviously related
with that of the $r=2$ and $r=3$ fields:
\be
\label{z}
\left.
\begin{array}{l} \displaystyle {\mu=-L=\sigma J^k, ~ ~
\tilde{G}^\mu =\Xi\delta^\mu_t, ~ ~ \Xi=\frac{1}{\sqrt{g_{00}}}
\left(\frac{\mu}{\sigma}\right)^{1/(2k)}},\\
\displaystyle{G=dB=d\left(\frac{\sqrt{3/2}}{M(K)}\right)\wedge
d\left( \frac{dL}{dJ}\tilde{G}\right)} \end{array} \right\}
\ee
({\em cf.} (\ref{x})). The function $M$ depends arbitrarily on
coordinates; thus one can choose its adequate form using the last
relation without coming into contradiction with the dynamical
Einstein--Euler equations.

When one describes a fluid in its proper basis, $u=J^{-1/2}\tilde
G=\theta^{(0)}$, the rotation of the fluid's co-moving reference
frame is defined as $\omega=\ast\left(\theta^{(0)}\wedge d
\theta^{(0)}\right)=J^{-1}\ast(\tilde G\wedge d\tilde G)$. Let us
assume $\theta^{(0)}=e^\alpha(dt+fd\phi)$ where $\alpha$ and $f$
are functions of coordinates (usually determined {\em via}
Einstein's equations), {\em cf.} the examples of metrics
considered in the next Sections, though in these examples are
treated Einstein--Maxwell fields and still not the perfect fluid
solutions. It is inevitable to conclude that the field theoretic
approach to perfect fluids automatically gives hints and even
concrete relations (often having a simple algebraic form) imposed
upon these and other functions characterizing the metric tensor
and the 2-form field, as well as the Machian one. This makes it
possible to substantially simplify the treatment of Einstein's
equations.

\section{A spherically symmetric dust-space-time
with NUT-like rotation}

We begin considering the following line element
$$
ds^2=\hbox{\large e}^{2\alpha(r)}[dt+l\cos\vartheta d\varphi]^2
-\hbox{\large e}^{2\beta(r)}dr^2
-r^2(d\vartheta^2+\sin^2\vartheta d\varphi^2),
$$
together with the orthonormal basis
$$
\theta^{(0)}=\hbox{\large e}^{\alpha}[dt+l\cos\vartheta d\varphi],~~~~
\theta^{(1)}=\hbox{\large e}^{\beta}dr,~~~~
\theta^{(2)}=rd\vartheta,~~~~
\theta^{(3)}=r\sin\vartheta d\varphi.
$$
Now we introduce the following 2-form, in order to satisfy (\ref{x}),
the 2-form should have the structure,
$$
B=F(r)\sin\vartheta d\vartheta\wedge d\varphi,
$$
to avoid a dependence of $\vartheta$ in $J$, we have introduced the
The respective field intensity being
$$
G=dB=\frac{F'(r)\hbox{\large e}^{-\beta}}{r^2}\theta^{(1)}\wedge
\theta^{(2)}\wedge\theta^{(3)},
$$
and
$$
\tilde G=*G=\frac{F'(r)\hbox{\large e}^{-\beta}}{r^2}\theta^{(0)}.
$$
The corresponding invariant
$$
J=\tilde G\cdot\tilde G=\left(\frac{F'}{r^2}\hbox{\large e}^
{-\beta}\right)^2.
$$
When we consider the dust case, $L=-\sigma\sqrt{J}$, the field equations
(\ref{x}) take the form
$$
d\left[-\frac{\sigma}{2}\hbox{\large e}^{\alpha}(dt+l\cos\vartheta d\varphi)\right]
=\sqrt\frac{2}{3}M(K)F(r)\sin\vartheta d\vartheta\wedge d\varphi.
$$
From them we conclude that $\alpha=0$ and
$$
M(K)=\sqrt\frac{3}{8}\frac{l}{\sigma F'}.
\label{M}
$$
With this conclusion Einstein's equations are
$$
G_{(0)(0)}=\frac{1}{r^2}\left(
\hbox{\large e}^{-2\beta}-1
-\frac{l^2}{4r^2}\right)
+\frac{1}{r}\left(\hbox{\large e}^{-2\beta}\right)'
-\frac{l^2}{2r^4}=
-\varkappa \mu
$$
$$
G_{(1)(1)}=-\frac{1}{r^2}\left(
\hbox{\large e}^{-2\beta}-1
-\frac{l^2}{4r^2}\right)=0,
$$
$$
G_{(2)(2)}=G_{(3)(3)}=
-\frac{1}{2r}\left(\hbox{\large e}^{-2\beta}\right)'
-\frac{l^2}{4r^4}
=0.
$$
Immediately we find that
$$
\hbox{\large e}^{-2\beta}=1
+\frac{l^2}{4r^2},~~~~~~\mu=\frac{l^2}{\varkappa r^4}.
$$
From the field theoretic description of the mass density
$$
\mu=\sigma\sqrt J=\frac{\sigma\sqrt{4r^2+l^2}}{2r^3}F',
$$
we find
$$
F(r)=-\frac{2l}{\sigma\varkappa}\ln\left
(\frac{l+\sqrt{4r^2+l^2}}{2r}\right)
$$
and form (\ref{M})
$$
M(K)=-\sqrt\frac{3}{32}\varkappa\left[
\ln\left
(\frac{l+\sqrt{4r^2+l^2}}{2r}\right)
\right]^{-1}.
$$
\subsubsection{The Solution}
Doing the change $l\rightarrow 2l$, the solution becomes
\begin{equation}
ds^2=[dt+2l\cos\vartheta\, d\varphi]^2
-\frac{r^2}{r^2+l^2}dr^2
-r^2(d\vartheta^2+\sin^2\vartheta\, d\varphi^2),
\label{polvo}
\end{equation}
with
$$
\mu=\frac{4l^2}{\varkappa r^4},~~~~
B=-\frac{4l}{\sigma\varkappa}\ln\left
(\frac{l+\sqrt{r^2+l^2}}{r}\right) \sin\vartheta\,d\vartheta\wedge
d\varphi
$$
and
$$
M(K)=-\sqrt\frac{3}{32}\varkappa\left[
\ln\left(\frac{l+\sqrt{r^2+l^2}}{r}\right)
\right]^{-1}.
$$
Note that if $l=0$, we arrive to the Minkowski space-time.

\subsubsection{Petrov type and isometries}
This solution if of Petrov type $D$, and it possesses an isometry
group $G_4$. Killing vectors are
$$
\xi_{[0]}=\partial_t,
$$
$$
\xi_{[1]}=\partial_\varphi,
$$
$$
\xi_{[2]}=2l\frac{\cos\varphi}{\sin\vartheta}\partial_t-\sin\varphi\,
\partial_\vartheta-\cot\vartheta \cos\varphi d\varphi,
$$
$$
\xi_{[3]}=-2l\frac{\sin\varphi}{\sin\vartheta}\partial_t-\cos\varphi\,
\partial_\vartheta-\cot\vartheta \sin\varphi d\varphi.
$$
They satisfy
$$
\left[\xi_{[0]},\xi_{[i]}\right]=0,~~~~
\left[\xi_{[i]},\xi_{[j]}\right]=\varepsilon_{ijk}\xi_{[k]},
$$
where $i$, $j$, $k$ = 1,2,3. We see the space-time has a
spherical-like symmetry.

\subsubsection{Closed timelike geodesics}
The angular coordinate $\varphi$ plays the r\^ole of a timelike
coordinate for some values of the other coordinates.
Now consider a motion with constant $t$, $r$ and $\vartheta$.

From geodesic equation we find the first integral
$$
(4l^2\cos^2\vartheta-r^2\sin^2\vartheta)
\left(\frac{d\varphi}{ds}\right)=\Lambda.
$$
Substituting it in the line element $ds^2$, we arrive to
$$
\frac{\Lambda^2}{4l^2\cos^2\vartheta-r^2\sin^2\vartheta}
=1.
$$
Thus, if $4l^2\cos^2\vartheta-r^2\sin^2\vartheta>0$,
we find that our solution accepts closed timelike geodesics for
$$
\Lambda=\sqrt{4l^2\cos^2\vartheta-r^2\sin^2\vartheta}.
$$
In the particular case $r=2l$, it is required that
$\vartheta<\pi/4$ \'o $\vartheta>3\pi/4$.

\subsubsection{The co-moving reference frame}
The co-moving reference frame with dust-particles is described by
the monad (dust 4-velocity, see [1]).
$$
\tau=\theta^{(0)}=dt+2l\cos\vartheta\,d\varphi.
$$
This reference frame is rotating
$$
\omega:=\frac{1}{2}*(\tau\wedge d\tau)=
-\frac{l}{r\sqrt{l^2+r^2}}dr,
$$
but has no acceleration, expansion, and shear (the rate of strain
tensor vanishes).

\subsubsection{Curvature}
We see that the curvature presents a strong singularity
$$
R_{(0)(2)(0)(2)}=R_{(0)(3)(0)(3)}=
R_{(1)(2)(1)(2)}=R_{(1)(3)(1)(3)}=-\frac{l^2}{r^4}
$$
$$
R_{(2)(3)(2)(3)}=-2\frac{l^2}{r^4}.
$$
$$
R_{(0)(1)(2)(3)}=2R_{(1)(2)(3)(0)}=-2R_{(1)(3)(2)(0)}=2\frac{l\sqrt{l^2+r^2}}{r^4}
$$

\vspace*{1.cm}

\noindent{\large\bf ACKNOWLEDGMENTS}

\vspace*{0.5cm}

\noindent I wish to thank Nikolai V. Mitskievich (Thesis
Supervisor) for helpful remarks and comments. I also wish 
to thank J\'ozsef Zsigrai for remarks made on this paper.
This paper contains a part of my Ph. D. Thesis, and the 
CONACyT-M\'exico scholarship grant No. 91290 is gratefully 
acknowledged.

\vspace*{1.cm}

\noindent{\large\bf REFERENCES}\footnotesize{

\vspace*{0.5cm}

\noindent 1. Luk\'acs B, Newman E T, Sparling G and Winicour J
(1983) A NUT-like solution with fluid matter {\it Gen. Rel Grav.} {\bf
15}, 567-579.}

\noindent 2. Mitskievich N V (1996). {\it Relativistic Physics
in Arbitrary Reference Frames}, gr-\phantom{aaa}qc/9606051.

\noindent 3. Mitskievich N V (1999). {\it Int. J. Theor. Phys.}
{\bf 38}, 997.

\noindent 4. Mitskievich N V (1999). {\it Gen. Rel Grav.} {\bf
31}, 713.

\noindent 5. Mitskievich N V (2003). {\it Rev. Mex. de
F{\'\i}sica} {\bf 49 Supl. 2}, 39; (2002). {\it Spacetimes,
\phantom{aaa}electromagnetism and fluids (a revision of
traditional concepts)}, gr-qc/0202032.

\noindent 6. Zsigrai J (2003) {\it Ellipsoidal shapes in general 
relativity: general definitions and an \phantom{aaaa} 
\phantom{a} application}, 
gr-qc/0301019.

\end{document}